\documentclass[sigconf]{acmart}
\AtBeginDocument{%
  \providecommand\BibTeX{{%
    \normalfont B\kern-0.5em{\scshape i\kern-0.25em b}\kern-0.8em\TeX}}}
\usepackage{xcolor}
\usepackage{subfig}
\usepackage{enumitem}

\copyrightyear{2025}
\acmYear{2025}
\setcopyright{cc}
\setcctype{by}
\acmConference[WWW '25]{Proceedings of the ACM Web Conference 2025}{April
28-May 2, 2025}{Sydney, NSW, Australia}
\acmBooktitle{Proceedings of the ACM Web Conference 2025 (WWW '25), April
28-May 2, 2025, Sydney, NSW, Australia}
\acmDOI{10.1145/3696410.3714865}
\acmISBN{979-8-4007-1274-6/25/04}

\begin{document}

\title{Causal Insights into Parler's Content Moderation Shift: \\ Effects on Toxicity and Factuality}

\author{Nihal Kumarswamy}
\orcid{0000-0002-0900-7634}
\authornote{Both authors contributed equally to the paper}

\affiliation{%
  \institution{The University of Texas at Arlington}
  \city{Arlington}
  \state{Texas}
  \country{USA}
}
\email{nihal.kumarswamy@mavs.uta.edu}

\author{Mohit Singhal}
\orcid{0000-0002-7423-9116}
\authornote{Corresponding author.}
\authornote{Work done at The University of Texas at Arlington}
\authornotemark[1]
\affiliation{%
  \institution{Northeastern University}
  \city{Boston}
  \state{Massachusetts}
  \country{USA}
}
\email{m.singhal@northeastern.edu}

\author{Shirin Nilizadeh}
\orcid{0000-0002-0539-3742}
\affiliation{%
  \institution{The University of Texas at Arlington}
  \city{Arlington}
  \state{Texas}
  \country{USA}
}
\email{shirin.nilizadeh@uta.edu}

\renewcommand{\shortauthors}{Nihal Kumarswamy, Mohit Singhal, \& Shirin Nilizadeh}

\begin{abstract}
Social media platforms employ various content moderation techniques to remove harmful, offensive, and toxic content, with moderation levels varying across platforms and evolving over time. Parler, a fringe platform popular among conservative users, initially had minimal moderation, promoting itself as a space for open discussion. However, in 2021, it was removed from the Apple and Google App Stores and suspended from Amazon Web Services due to inadequate moderation of harmful content. After a month-long suspension, Parler returned with stricter guidelines, offering a unique opportunity to study the impact of platform-wide policy changes on user behavior and content outcomes.
In this paper, we analyzed Parler data to assess the causal associations of these moderation changes on content toxicity and factuality. Using a longitudinal dataset of 17M posts from 432K users, who were active both before and after replatforming, we employed quasi-experimental analysis, controlling for confounding factors. 
We introduced a novel approach by using data from another social media platform, Twitter, to account for a critical confounding factor: offline events. This allowed us to isolate the effects of Parler's replatforming policies from external real-world influences. 
Our findings demonstrate that Parler's moderation changes are causally associated with a significant reduction in all forms of toxicity ($p<0.001$). 
Additionally, we observed an increase in the factuality of the news sites shared and a reduction in the number of conspiracy/ pseudoscience sources.

\end{abstract}



\maketitle
\begin{center}
\large
\textcolor{red}{\textbf{This paper has been accepted at WWW 2025, please cite accordingly.}
}\end{center}

\section{Introduction} 
Social media has become a powerful tool that reflects the best and worst aspects of human communication. 
On one hand, they allow individuals to freely express opinions, engage in interpersonal communication, and learn about new trends and stories. 
On the other hand, they have also become fertile grounds for several forms of abuse, harassment, and the dissemination of  misinformation~\cite{fake_news_2021,khan_2019,anderson2016toxic,singhal2021prevelance}. 
Social media platforms, hence, continue to adopt and evolve their content moderation techniques and policies to address these issues while trying to respect freedom of speech and promote a healthier online environment. 

Social media platforms, however, do not follow unified methods and policies for content moderation~\cite{singhal2022sok}. 
While some social media platforms adopt more stringent content moderation rules, others, like Parler, pursue a laissez-faire approach. Parler, launched in 2018, adhered to this hands-off moderation philosophy, contending that it promoted richer discussions and protected users' freedom of speech~\cite{rothschild_2021}. This was until January 6th, 2021, when Parler gained much notoriety for being home to several groups and protesters who stormed Capitol Hill~\cite{groeger2017parler,parler-role}. 
Subsequently, due to its content moderation policies and concerns about the spread of harmful or extremist content, Parler faced significant consequences. It was not only terminated by its cloud service provider, Amazon Web Services but also removed from major app distribution platforms, including the App Store and the Google Play store~\cite{fung_2021}. 

For Paler to return, it had to enact substantial revisions to its \emph{hate speech policies}.\footnote{Example of changes in Parler CG: \url{https://tinyurl.com/yda6pfmj}.} This included a complete removal of the ability for users on iOS devices to access objectionable and Not Safe for Work (NSFW) content. 
As a result, Parler's updated policies introduced more stringent moderation policies aimed at curbing hate speech on the platform~\cite{lerman_2021}. 
While prior studies have focused on the impact of dealtforming a small subset of users or specific communities~\cite{rogers2020deplatforming,rauchfleisch2021deplatforming,ali2021understanding,jhaver2021evaluating,ribeiro2020does,cima2024great}, or have examined how content moderation affects the activities of problematic users~\cite{trujillo2022make,ali2021understanding}, our work evaluates the impact of new platform-wide policy changes on the ``within-platform'' dynamics of all users who were active during both the pre- and post-policy periods. 
This inclusive approach offers a comprehensive perspective on the broader ecosystem, moving beyond the limited focus on dealtformed users or specific audiences. Furthermore, while prior research has predominantly examined hard content moderation measures—such as suspensions or removals—our study shifts the focus to \emph{replatforming}, a distinct scenario involving the reinstatement of a platform accompanied by a series of progressive policy changes.
In particular, we investigated two research questions: 

\textbf{RQ1:}~Did changes to Parler's content moderation guidelines had any significant impact on the user-generated content?

\textbf{RQ2:}~How have Parler's content moderation revisions changed its existing users' characteristics?

To assess these effects, we conducted a quasi-experimental analysis, called Difference-in-Difference (DiD)~\cite{abadie2005semiparametric}, monitoring user posts for toxic content, insults, identity attacks, profanity, and threats. In addition, we explored shifts in users' characteristics and conversation topics, and quantified the presence of biased posts and posts with non-factual links, utilizing data sourced from Media Bias Fact Check (MBFC)~\cite{mbfc_about_2022}. 
We used the data from Aliapoulios et al.~\cite{aliapoulios2021early} as the seed dataset (we call this dataset a \emph{pre-policy change} dataset), and we tried to collect the posts for the same sample of 4M users. Developing our custom build crawler, we collected about 17M parleys of a subset of 432K users who were active from February 2021 to January 2022. We labeled our dataset as a \emph{post-policy policy change} dataset. To the best of our knowledge, ours is the first dataset that was collected after Parler came back online. 
To measure the effect of Parler's content moderation changes, we used the Difference-in-Difference (DiD) regression analysis, which is arguably one of the strongest and widely used quasi-experimental methods in causal inference~\cite{horta2023deplatforming,han2019causal,crown2014propensity,fredriksson2019impact}. This analysis helped us understand how and if the outcomes, e.g., the number of toxic posts, have changed after Parler changed its moderation guidelines. 

In DiD analysis, to account for the potential influence of offline events, such as social or political unrest, on platform toxicity, we used data from another social media platform—Twitter—as a control group. Our reasoning is that offline events, such as elections, would likely increase toxicity across all platforms, regardless of their content moderation policies. For example, even with strict moderation, Twitter would probably experience heightened toxicity during periods of unrest compared to calmer times. To minimize bias, we analyzed a random sample of Twitter discussions, rather than specifically targeting far-right conversations or incorporating data from other fringe platforms. 

Thus, this paper has the following contributions and findings: 
\begin{enumerate}
\item Our work demonstrates how the effectiveness of content moderation policies can be evaluated through data-driven analysis using platform-specific data, in this case, Parler, before and after its moderation policy changes.

\item We collected the first-ever post-replatforming dataset from Parler.

\item Using the Difference-in-Differences (DiD) approach, we found that Parler was effective in reducing various types of toxicity.

\item While most related studies fail to account for external offline events that may influence user activity or toxicity, we used trends from a random sample of Twitter data as a control group in our DiD analysis to isolate the effects of the replatforming policies.

\item Our findings showed an increase in both follower and following counts, along with a rise in users with verified and gold badges. This suggests potential growth in Parler's user base, as well as the continued presence of older users who were active before the moderation policy changes.

\item We observed an improvement in factuality and credibility scores from the pre-moderation dataset to the post-moderation dataset. Additionally, there was a reduction in the sharing of conspiracy and pseudoscience source links. However, an increase was noted in the sharing of questionable source links in the post-moderation dataset. 

\end{enumerate}

\section{Related Works}
\textbf{Fringe Communities: } Over the past few years, scholars have extensively studied various fringe platforms such as Gab and 4chan~\cite{hine2017kek,bernstein20114chan,zannettou2018gab,sipka2022comparing,jasser2023welcome}. In contrast, Parler is a relatively younger platform, resulting in fewer studies focusing on collecting data or establishing frameworks for data collection from Parler~\cite{aliapoulios2021large,prabhu2021capitol}.
Some studies have compared topics of discussion on Parler and Twitter~\cite{prabhu2021capitol,sipka2022comparing}, most focusing on the presence or prevalence of a single topic. Hitkul et al.\cite{prabhu2021capitol} examined the Capitol riots—a pivotal event in Parler's history—to compare discussions on Parler and Twitter. Other works have analyzed language use on Parler across various topics, such as QAnon content~\cite{sipka2022comparing,bar2023finding} and COVID-19 vaccines~\cite{baines2021scamdemic}. Our work differs in that we specifically study changes within Parler itself, focusing on how users reacted to the platform’s temporary hiatus.

\textbf{Studies about Deplatforming: } 
All existing studies on deplatforming examine a defined group of deplatformed users and their audiences. Some studies focus on a small number of users (e.g., three users~\cite{jhaver2021evaluating}), while others analyze users from specific subreddits (e.g., two subreddits~\cite{horta2021platform,trujillo2022make} or 15 subreddits~\cite{cima2024great}) or certain Telegram channels~\cite{rogers2020deplatforming}. Since these users are already deplatformed, these studies focus on how their behavior changed after migrating to a new platform. For example, they examine whether activity levels were affected or if there was any change in the use of hate speech~\cite{ali2021understanding,horta2023deplatforming,rauchfleisch2021deplatforming,horta2021platform,rogers2020deplatforming,russo2023spillover}.  
Some studies also track the behavior of audiences across the original and new platforms, investigating questions such as whether followers also migrated to the second platform~\cite{rauchfleisch2021deplatforming,rogers2020deplatforming,russo2023spillover} and, if so, whether they became more extreme in their language~\cite{ali2021understanding,horta2023deplatforming,horta2021platform}. 
A common finding across these studies is that deplatforming significantly reduces the reach of deplatformed users; however, it also tends to intensify hateful and toxic rhetoric within their new online spaces. 

While most of these studies examine user behavior across platforms, only a few focus on ``within-platform'' dynamics~\cite{horta2021platform,trujillo2022make}. 
These works investigate the deplatforming of a small set of subreddits on Reddit, analyzing whether and how the deplatformed users migrated to other subreddits within the same platform. In other words, even when focusing on ``within-platform'' effects, these studies primarily aim to understand the behavior of a set of deplatformed users and their audiences. 

\textbf{Hate Speech Detection and Classification: }
Empirical work on toxicity has employed machine learning-based detection algorithms to identify and classify offensive language, hate speech, and cyberbullying~\cite{davidson2017automated,pitsilis2018detecting}. Features including lexical properties, such as n-gram features~\cite{nobata2016abusive}, character n-gram features~\cite{mehdad2016characters}, character n-gram, demographic and geographic features~\cite{waseem2016hateful}, sentiment scores~\cite{dinakar2012common,sood2012profanity}, average word and paragraph embeddings~\cite{nobata2016abusive,djuric2015hate}, and linguistic, psychological, and affective features inferred using an open vocabulary approach~\cite{elsherief2018hate} have been used to detect hate speech.
Google's Perspective API\cite{jigsaw2018perspective} has been widely utilized in prior studies\cite{salehabadi2022,zannettou2020measuring,grondahl2018all,elsherief2018hate,saveski2021structure,papasavva2020raiders,aleksandric2022twitter,hickey2023auditing} to assess the toxicity of online content. Developed by Jigsaw, the API employs ML models to assign toxicity scores to text based on various attributes, such as insult, threat, identity attack, and profanity. Despite its widespread adoption, studies have also critiqued its biases, particularly its tendency to over-penalize certain linguistic styles and dialects, raising concerns about fairness and reliability in automated moderation~\cite{hosseini2017deceiving}.

\textbf{Media Bias Fact Check (MBFC): }
MBFC is widely used to assess the credibility and factuality of news sources for downstream analysis~\cite{main2018rise,heydari2019youtube,starbird2017examining,darwish2017trump,nelimarkka2018social,etta2022comparing,bovet2019influence,cinelli2020covid,cinelli2021echo,weld2021political} and serves as ground truth for prediction tasks~\cite{dinkov2019predicting,stefanov2020predicting,patricia2019link,bozarth2020higher,gruppi2021nela}.
Gruppi et al.~\cite{gruppi} used MBFC service to label websites and the tweets pertaining to COVID-19 and 2020 Presidential elections embedded inside these articles. 
Weld et al.~\cite{weld2021political} analyzed more than 550M links spanning 4 years on Reddit using MBFC. 

\section{Methodology}
To address our research questions, we first utilized the existing Parler data collected before the policy change~\cite{aliapoulios2021early} and then developed a framework for collecting longitudinal data following the policy change. Second, we utilized Google's perspective API~\cite{jigsaw2018perspective} on all the posts to measure various types of toxicity. Third, we analyzed all the links provided in the posts for bias and factuality, using MBFC services.  
Fourth, we employed Difference-in-difference (DiD) model~\cite{abadie2005semiparametric}, a quasi-experimental approach, to measure the causal associations of Parler's content moderation change on \textit{toxicity attributes}. 
In this DiD analysis, we proposed using data from Twitter as a control group to account for a key confounding variable: offline events. Major and contentious events, such as the U.S. presidential election, often drive significant spikes in online activity and can influence toxicity levels across social media platforms. We hypothesize that similar trends in toxicity may be observed on Twitter, which did not implement any policy changes during the same period. By comparing these trends with those observed on Parler, we could assess whether the observed shifts in toxicity levels are unique to Parler's content moderation adjustments or part of broader online dynamics. Therefore, we gathered a sample of Twitter data from the same timeframe and analyzed the trends before and after Parler's policy changes. This approach allows us to isolate the effects of Parler’s policy shifts and draw more robust conclusions about their impact on toxicity.

Finally, we examined factors such as the number of followers, following, badge changes, the topics of conversation, and any shifts in the \textit{bias} and \textit{credibility} of the URLs being shared. 

\subsection{Data Collection}
\label{datacollection}

\textbf{Pre Policy Change Dataset: }
Aliapoulios et al.~\cite{aliapoulios2021early} developed a data collection tool to gather user information from Parler, capturing data from nearly all active users at the time. The study collected user information from over 13.25M users and randomly selected 4M users for further analysis. For these users, approximately 99M posts (or ``parleys'') and 85M comments were gathered from August 1, 2018, to January 11, 2021. In our study, we refer to this dataset as the \emph{pre-policy change} dataset. We used these 4M users as a seed dataset to collect data following the policy changes. 

\textbf{Post Policy Change Parler Dataset: }
We obtained the list of 4M users provided in the pre-policy change dataset for which the authors collected posts and comments~\cite{aliapoulios2021large} and used our custom-build framework to get the content posted by the same users. 
Using our framework, we collected information about the post body, any URLs posted, a URL to the location of any media posted, the date posted, the number of echoes, and other metadata, such as the badges of the poster. 
Authors in~\cite{aliapoulios2021early} obtained the metadata of 13.25M users, hence we also tried to collect the metadata of these users. 

\textbf{Post Policy Change Dataset Statistics:}  
From the 4M users, we could collect 17,389,610 parleys from 432,654 active users. Our dataset consists of parleys from February 1st, 2021 to January 15th, 2022. We used the \textit{/pages/feed} endpoint, which returns the parleys (posts) posted by a specific user using their username. 
Note that, this endpoint is different from the endpoint that is used to collect the 13.25M users' metadata, and hence we were only able to obtain 432K users' parleys.
Several users from the initial seed dataset of 4M were no longer active. Manually checking these accounts we found that they had either deleted their accounts, or changed their usernames, or did not post any parley after Parler's return, or switched to private accounts. Note, we did not include any users' post if their account was private. 
We label them as \emph{missing} users. Even though we are unsure if these users were suspended by Parler or they decided to leave Parler, we nevertheless analyzed and compared these users with those that remained active. Since we only collected posts from 432,654 users, we acknowledge that certain trends and analyses conducted might not be accurately reflected on the platform. However, as of January 2022, months after returning to the Apple app store, Parler disclosed that they estimate to have around 700K to 1M active users~\cite{parler-active1}. This ensures that we have collected a significant part of the data. 

These parleys (17M) consisted of users posting around 9M links and plain text in the body. A majority of these posts were primary posts that had no parent. If a parley is an original parley and is not an echo of another parley, it is known as a primary post with no parent. We collected the full-text body, a URL if a link was shared, the title of the parley, the date of creation, flags for trolling, sensitive and self-reported, an upvoted flag, a counter of echoes and likes. We noticed that Parler has a trolling flag, which might be set manually by moderators or automatically by the platform. 

We also tried to collect profile information for 13.25M users from the pre-policy change dataset. We used the \textit{pages/profile/view} endpoint, that returns the metadata of the user. We found that 12,497,131 of these users still had a valid Parler account, so we could collect metadata for these users. For a vast majority of the accounts, Parler returned the number of followers, the number of following, status (account available or deleted), the number of and types of badges given to the user, a description of all badges available on Parler at the time of collection, date of parley creation, whether the account is private or public, and also whether the account is being followed by or a follower of the user logged in. A minority of profiles have one or more of these fields missing due to changes on the Parler platform from when the user created the account and the time of data collection.

\textbf{Twitter Data Collection: } 
For a comparative baseline, we gathered a sample of Twitter data from the same timeframe and analyzed the trends before and after Parler's policy changes. 
To collect the data, we used Twitter V2 API~\cite{twitterapi} and collected posts daily using the exact timeline of the datasets.
To circumvent the API restrictions, we collected 27K posts daily and restricted the posts to English. After the data collection was done, we were able to collect 24.16M posts for a pre-policy timeline and 9.69M for a post-policy timeline.

\textbf{Ethical consideration.} 
We only gathered posts from Parler profiles set to public and did not attempt to access private accounts. 
We used the same backend APIs that a user browser would request data from. We only obtained the random sample from Twitter API and did not collect any metadata information about the users whose profiles were set as private.

\subsection{Measuring Toxicity Scores} 
We utilized the Google Perspective API, which is a state-of-the-art toxicity detection tool~\cite{jigsaw2018perspective}. 
This AI-based tool investigates the  provided text and assigns a score between 0 to 1, with a higher score indicating more severity for a particular attribute. 
We obtained the following attributes: \emph{Severe toxicity}, \emph{Profanity}, \emph{Identity Attacks}, \textit{Threats}, \textit{Insults}, \textit{Toxicity}. For our study, we collected the likelihood scores for each attribute. While collecting the scores, English was used as the default language for all posts since previous studies showed us that a large majority of Parler's userbase was using English as their language of choice to communicate with other Parler users~\cite{aliapoulios2021early}. Before sending the posts to Perspective API, we pre-processed the posts by removing URLs, hashtags, etc. as these can lead to wrong scores or errors computing the scores by the API.
We also pre-processed our Twitter dataset and obtained the scores via Perspective API. 
While we acknowledge that the Perspective API has limitations in detecting toxicity~\cite{hosseini2017deceiving}, prior research has demonstrated its effectiveness in identifying various forms of toxic language in generated text~\cite{ovalle2023m}. Additionally, several studies have successfully used the Perspective API for toxicity detection~\cite{aleksandric2022twitter,elsherief2018hate,saveski2021structure,hickey2023auditing}. To validate the API's performance, two independent coders labeled 200 randomly selected Parleys. The inter-coder reliability, measured by the Cohen's Kappa score, was 0.7, indicating substantial agreement with the Perspective API's toxicity assessments. 

\subsection{Causal Inference}
We employed a causal inference strategy known as the Difference-in-Differences (DiD) model~\cite{abadie2005semiparametric} to measure the impact of Parler's content moderation changes. In our DiD analysis, we assess the causal effect of our dependent variable—toxicity attributes—over time using regression. This is done by comparing two groups: the treatment group (i.e., Parler, where changes in content moderation policies occurred after its return) and the control group (i.e., Twitter, where no such policy changes took place). It is crucial to account for time in the regression; otherwise, we risk misinterpreting a consistent trend (increasing or decreasing) as a treatment effect when simply comparing averages before and after the treatment~\cite{bernal2017interrupted}. 

To have a balanced dataset, and since our post-moderation dataset spans approximately 11 months, we filtered the pre-moderation dataset for 11 months (i.e., February 2020 to January 2021). We also employed the same step for our Twitter dataset. After filtering the data, we clustered the data points based on the day the parley or the tweet was posted. After clustering the data per day, we set the \emph{Perspective Score} for all the tweets' or parleys' as 0 if they were below 0.5, values above or equal to 0.5 and we kept the absolute value. We choose a threshold of 0.5, because prior research has used this threshold to distinguish if a post is toxic or not~\cite{aleksandric2022twitter,saveski2021structure}. We then averaged the scores per day to get a final score that we passed to our DiD regression model. To check the robustness of our method, we ran a regression model, where we did not use a threshold, and obtained the same results as we obtained when using the threshold, hence our model is robust. 

To estimate the effect of the content moderation changes, we employ a linear regression model to estimate the impact ($\delta$) of these policy adjustments following Parler's return: 
\begin{equation}
\label{eq}
Y = \beta_1 T + \beta_2 P + \delta TP + \epsilon
\end{equation}

In this model, $Y$ represents the toxicity score; $T$ is a binary variable indicating the treatment group (=1) and control group (=0); and $P$ is a binary variable indicating whether the observation was collected before (=0) or after (=1) the treatment. We estimate the coefficient $\delta$, which corresponds to the interaction between the variables $T$ and $P$, using Ordinary Least Squares (OLS) to obtain the average treatment effect. $\beta_1$ captures the difference between the treatment and control groups prior to the changes in Parler's content moderation guidelines, $\beta_2$ reflects the change in the outcome over time for the control group (i.e., Twitter, post-treatment), and $\delta$ represents the effect of Parler's content moderation policy changes on toxicity levels. 

We chose the DiD method over Interrupted Time Series (ITS) analysis because DiD is widely recognized in the econometrics and causal inference community for handling quasi-experimental interventions~\cite{berger2020tarp,yang2022effects}. Additionally, DiD provides a single causal estimate (i.e., $\delta$), simplifying the interpretation of results, whereas ITS generates six separate estimates (three for Parler and three for Twitter), making the interpretation more complex. 

\subsection{Examining Changes in User Characteristics and Content}
To answer RQ2 and understand if Parler's moderation change had an impact on its user base beyond users' speech, we performed additional analyses on factors such as the number of followers, following, badge changes, the topics of conversation, and any shifts in the \textit{bias} and \textit{credibility} of the URLs being shared. 

\subsubsection{User Characteristics} 
We extracted \emph{following} and \emph{followers} counts from both datasets to understand if any of these metrics have changed significantly after the moderation policy changes. Since these variables are not captured in time, and we have two different distributions, we cannot perform DiD regression analysis. 
The resulting values did not form a normal distribution, so we used the Mann-Whitney test. 
We also analyzed the number of badges assigned to each user in the pre- and post-moderation change datasets.

\subsubsection{Content Analysis}
We used textual data collected from Parleys  to investigate what users were discussing in both the pre- and post-policy change datasets. To identify the most popular topics, we applied the Latent Dirichlet Allocation (LDA) topic modeling technique~\cite{blei2003latent}. Before running LDA, we preprocessed the text by removing all URLs, Unicode characters, and stopwords. We also used a stopword corpus from the Natural Language Toolkit (NLTK) to filter out common words from our dataset.

\subsubsection{Assessing Bias and Factuality}  
We examined the links shared in Parleys to identify trends and align them with the rhetoric of online communities, providing a clearer understanding of the changes. To extract external links, we examined every Parley in both the pre- and post-policy change datasets for valid URLs. We then extracted the top-level domain names from each URL and recorded the frequency of each domain's occurrence to measure website popularity in each dataset.
Then, we analyzed these links using the Media Bias Fact Check (MBFC) service~\cite{mbfc_about_2022}. 
MBFC is an independent organization that uses volunteer and paid contributors to rate and store information about news websites~\cite{mbfc_about_2022}.
MBFC can be used to measure the factuality of the URL, the presence of any bias, the country of origin, and the presence of conspiracy or pseudoscience, questionable sources, and pro-science sources. 
We used a list of links shared from both of our datasets to obtain labels for:

\begin{itemize}[leftmargin=*]
    \item[] \textbf{Factuality}: Referred to as how factual a website is. Scored between 0-5, where a score of 0 means that a website is not factual and a five is very factual. MBFC defines that for a website to be very factual and get a score of 5, it should pass its fact-checking test as well as make sure that critical information is not omitted. 
    \item[] \textbf{Bias}: MBFC assigns a bias rating of Extreme left, left, left-center, least biased, right-center, right, and extreme right. To assign a bias rating to a website, MBFC contributors check the website's stance on American issues, which divides left-biased websites from right-biased websites~\cite{mbfcbias_2021}. 
    \item[] \textbf{Presence of conspiracy-pseudoscience}: Websites that publish unverified information related to known conspiracies such as the New World Order, Illuminati, False flags, aliens, anti-vaccine, etc.
    \item[] \textbf{Usage of questionable sources}: MBFC defines this as \emph{a questionable source exhibits any of the following: extreme bias, overt propaganda, poor or no sourcing to credible information, a complete lack of transparency, and/or is fake news. Fake News is the deliberate attempt to publish hoaxes and/or disinformation for profit or influence}.
\end{itemize}

\section{Results}
\label{perspective} 

\subsection{Impact of Stricter Content Moderation }
 
\begin{figure*}[t]
\centering 
\subfloat[Identity Attack] {\includegraphics[width=0.33\textwidth]{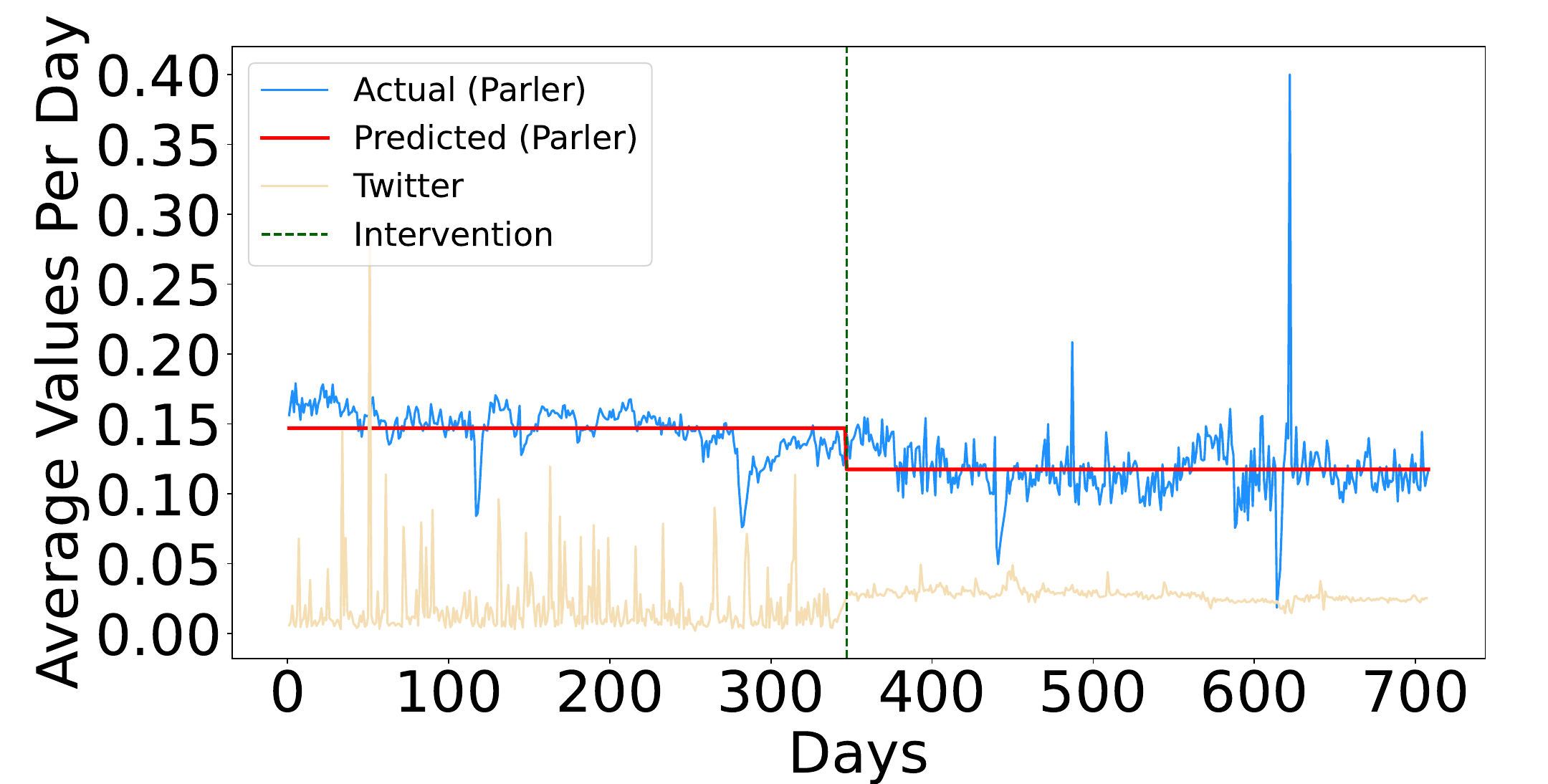}\label{fig:identity_attack}}
\subfloat[Insult]{\includegraphics[width=0.33\textwidth]{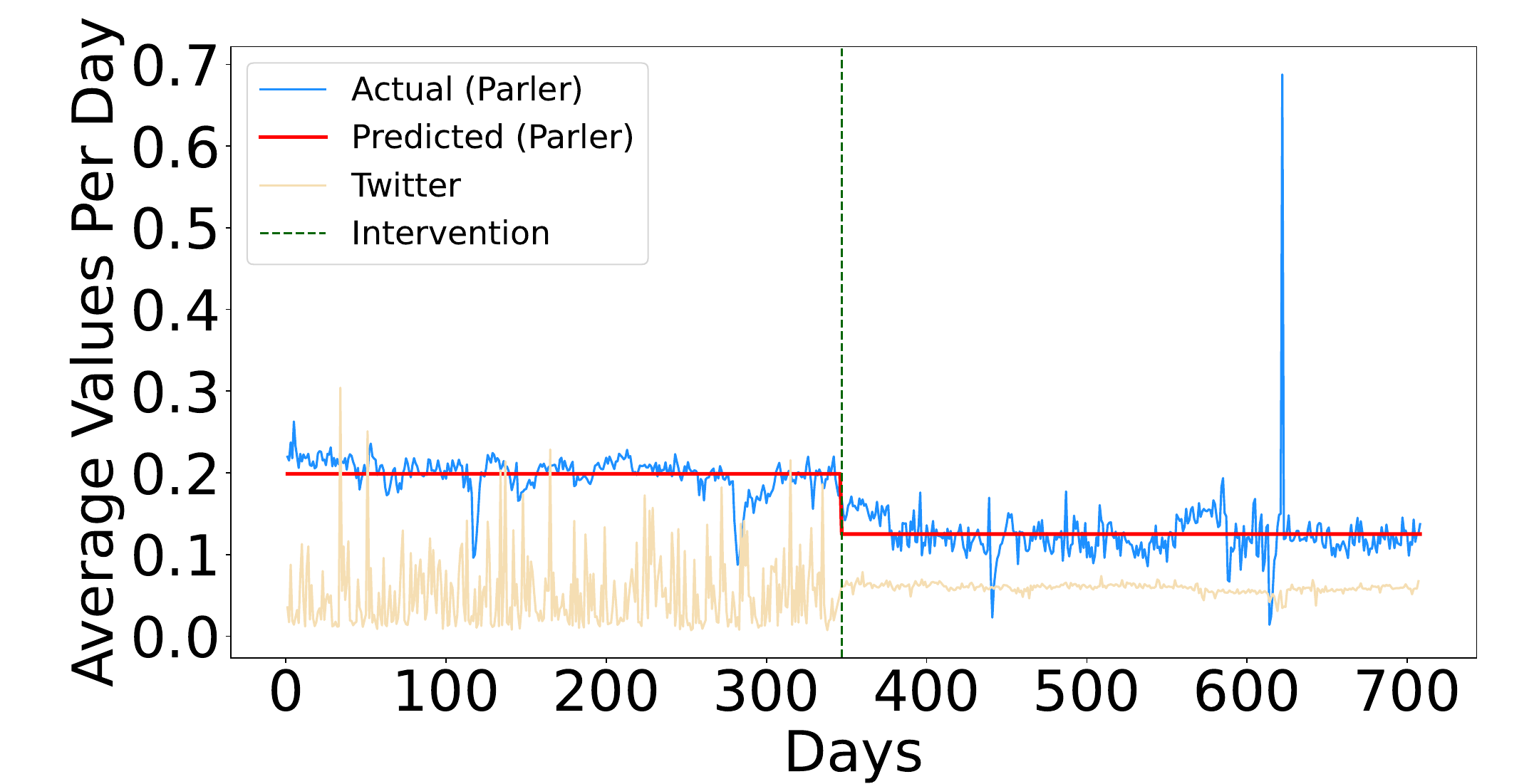} \label{fig:insult}} 
\subfloat[Profanity]{\includegraphics[width=0.33\textwidth]{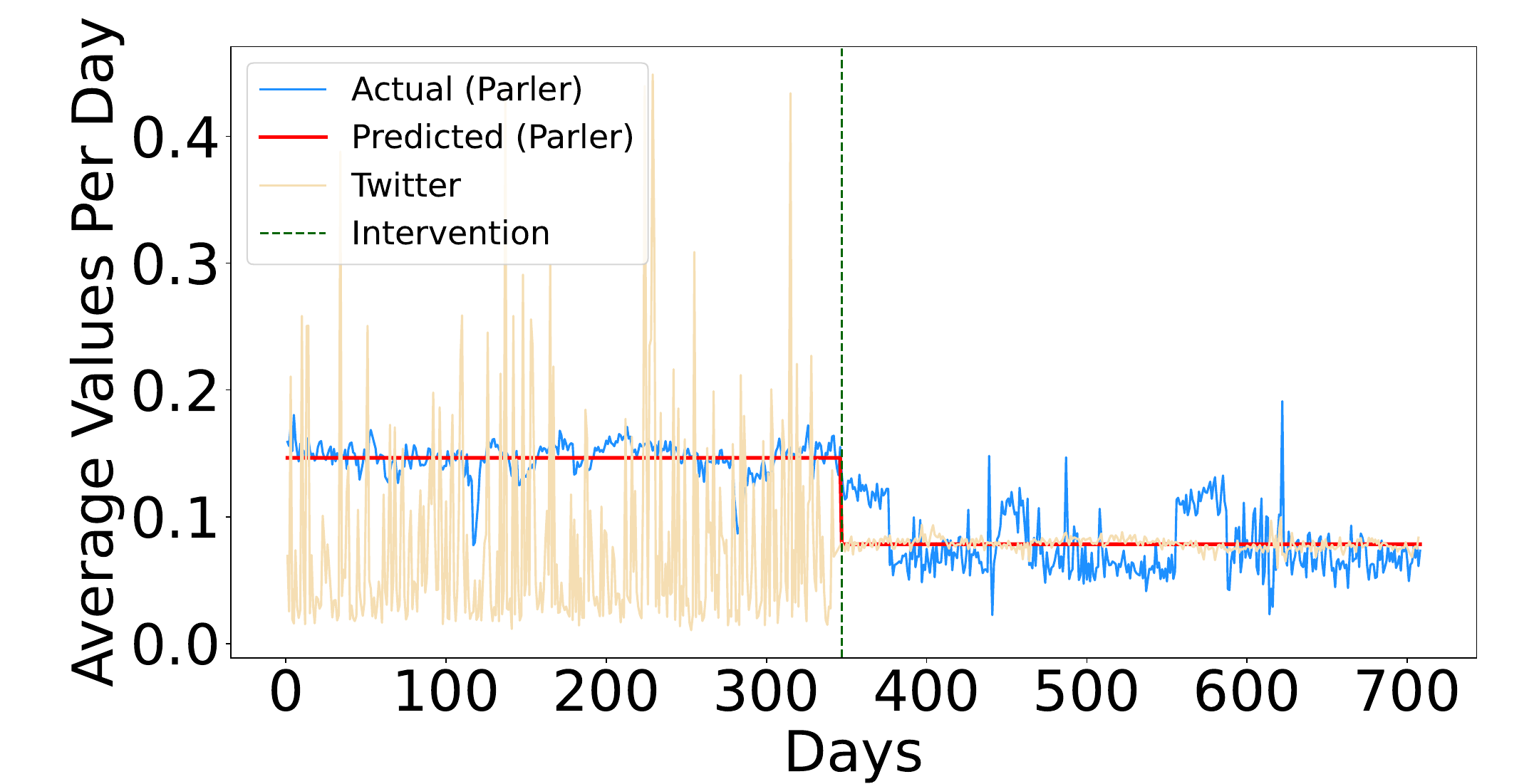} \label{fig:profanity}} \\
\subfloat[Severe Toxicity]{\includegraphics[width=0.33\textwidth]{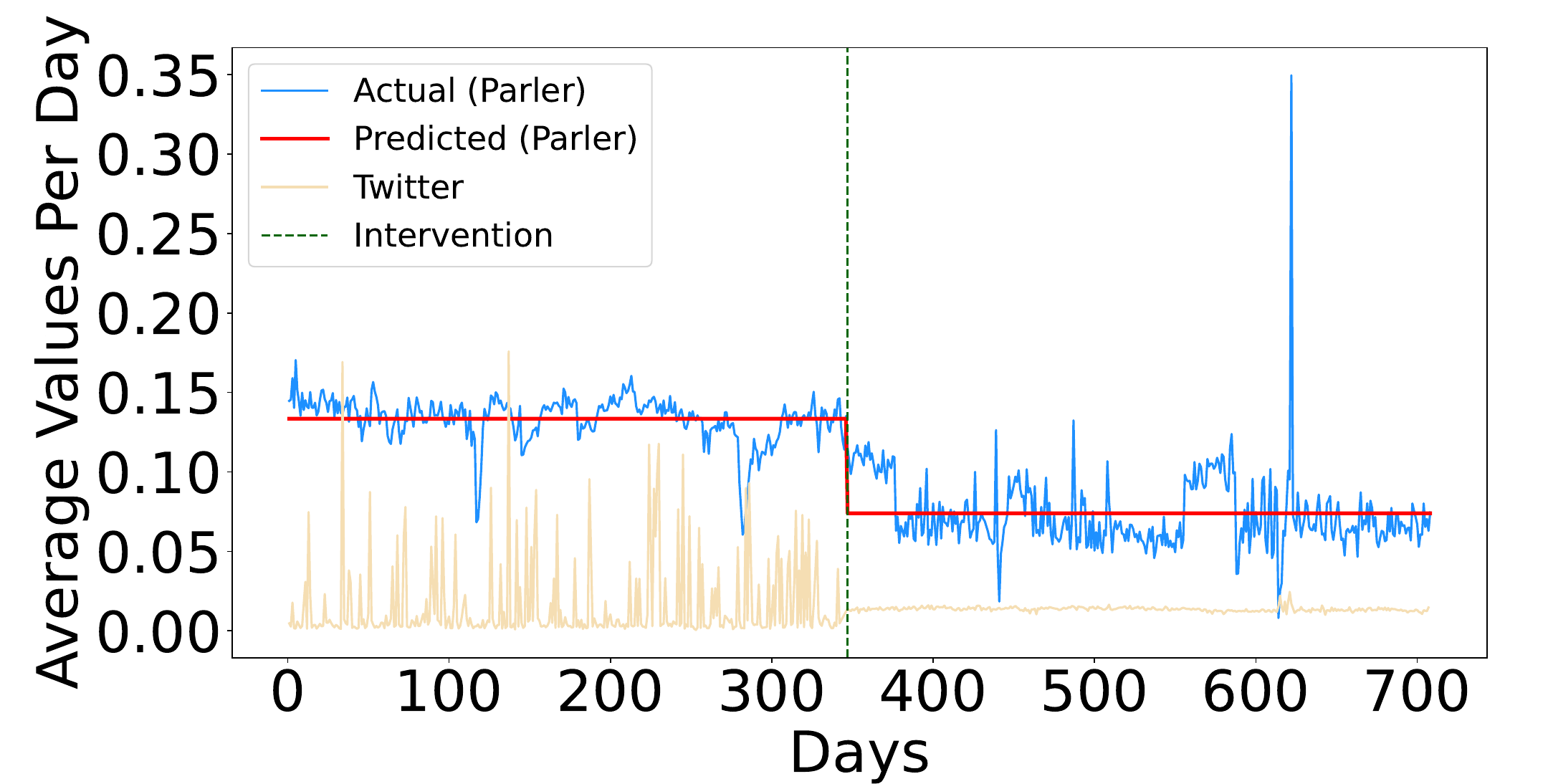} \label{fig:severe_toxicity}}
\subfloat[Threat]
{\includegraphics[width=0.33\textwidth]{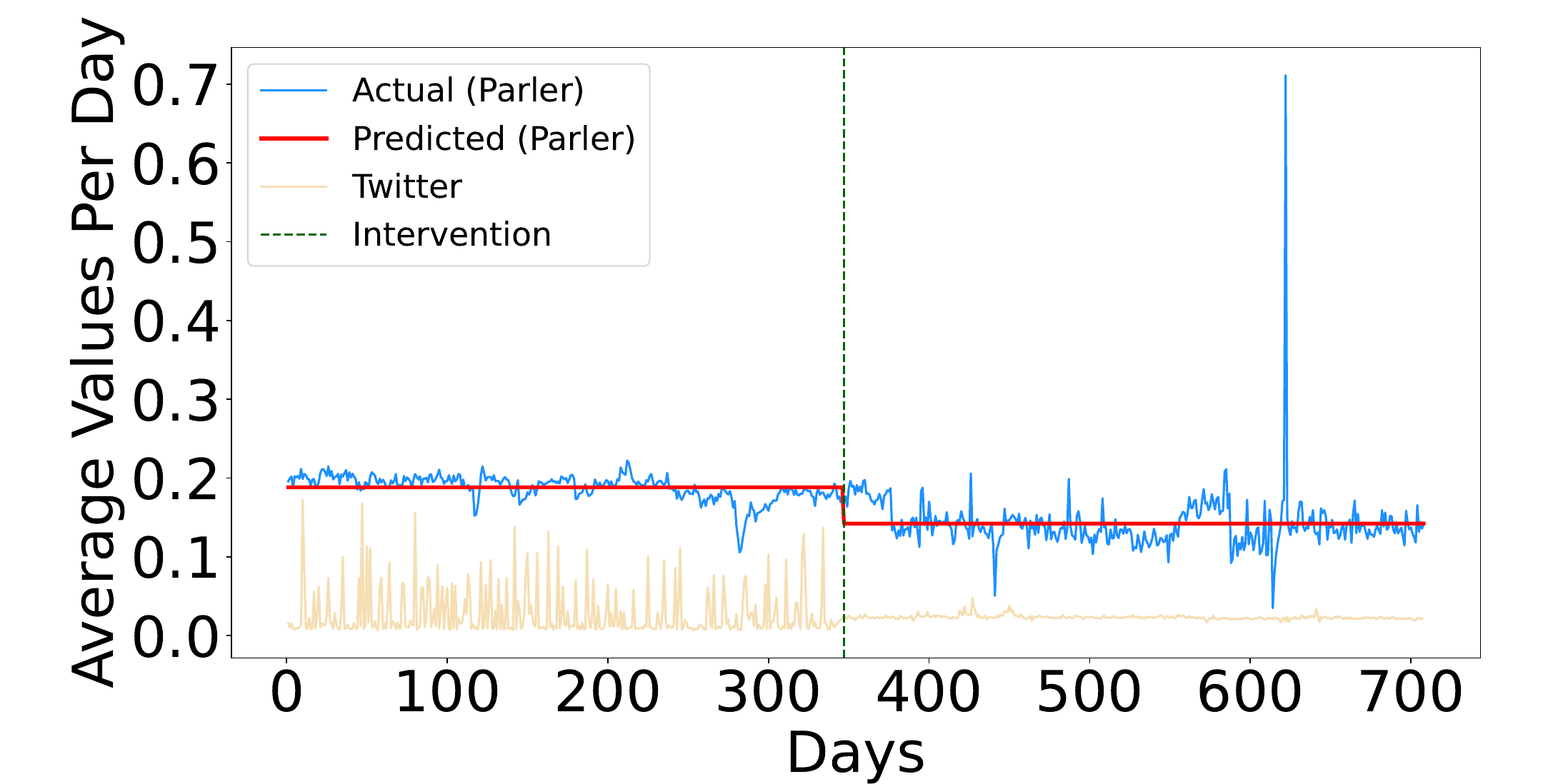} \label{fig:threat}} 
\subfloat[Toxicity]{\includegraphics[width=0.33\textwidth]{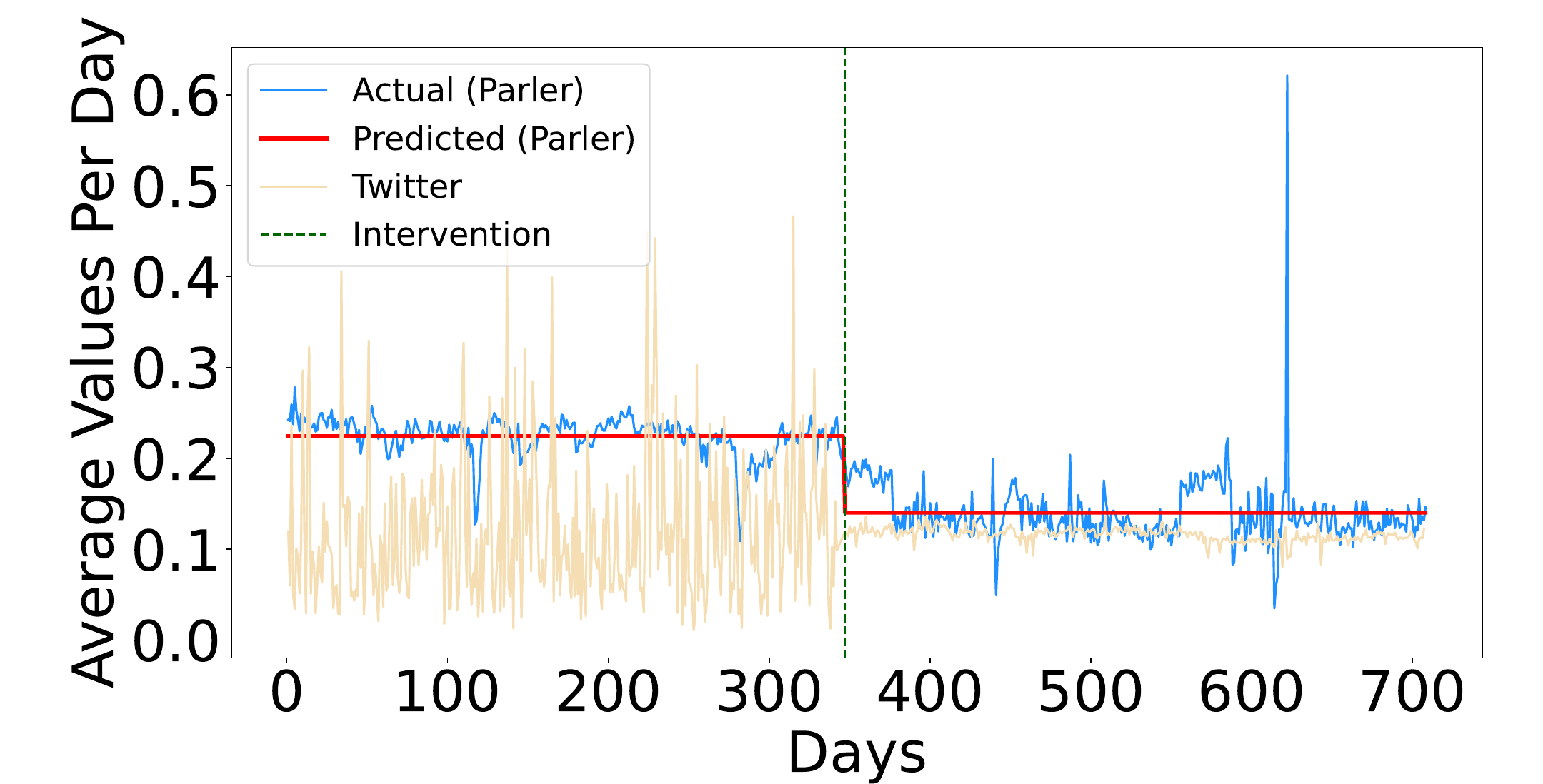} \label{fig:toxicity}} 
\caption{Difference in Difference (DiD) plots for Perspective Attributes. X-axis denotes the days, and y-axis denotes the average Perspective API scores.}  
\label{perspective-hist-i}
\end{figure*}

Table~\ref{stat-linearits} presents the results of our DiD analysis. We observe a causal association between Parler's return online and the implementation of changes to its content moderation guidelines, leading to significant decreases in \textit{Toxicity}, \textit{Severe Toxicity}, \textit{Profanity}, \textit{Threat}, \textit{Insult}, and \textit{Identity Attack} ($p < 0.001$). Additionally, our \emph{Treatment} variable, which indicates whether Parler and Twitter differed in their moderation effectiveness for each dependent variable prior to the policy changes, shows that, on average, Parler had higher levels of \textit{Toxicity}, \textit{Severe Toxicity}, \textit{Profanity}, \textit{Threat}, \textit{Insult}, and \textit{Identity Attack} compared to Twitter ($p < 0.001$). 
Interestingly, our \emph{Post Treatment} variable shows that Twitter users' \textit{Threat} posts also decreased ($p < 0.05$). This suggests that there could have been a general trend of reduced threatening posts, which could potentially challenge the findings for \textit{Threat} on Parler ($\delta$). However, when examining \textit{Insult} and \textit{Identity Attack}, we observed a statistically significant increase on Twitter ($p < 0.001$), while the opposite trend was observed for Parler ($\delta$). This indicates that despite the general decrease in \textit{Threat} posts, the changes to Parler's content moderation guidelines had a positive effect overall, leading to a reduction in abusive content on the platform. 
Additionally, Table~\ref{stat-linearits} presents the confidence intervals for all our variables. As observed, for all our \emph{dependent variables}, 95\% of the time we would expect the effect of \emph{DiD} ($\delta$) on the various dependent variables to fall within the respective lower and upper bounds. This further demonstrates that the differences in the proposed changes are statistically significant, particularly when examining Parler pre- and post-moderation changes. 

To further disentangle these findings, we plotted the results from our DiD model in Figure~\ref{perspective-hist-i}. The red solid line represents $\delta$ (i.e., the DiD estimate), while the green line indicates when Parler changed its content moderation guidelines and returned after its hiatus. As observed, across all attributes, the model reveals a statistically significant decrease, with the red line ($\delta$) dropping after the intervention in Parler. Notably, we can visually identify that \textit{Profanity} (Fig.\ref{fig:profanity}), \textit{Severe Toxicity} (Fig.\ref{fig:severe_toxicity}), and \textit{Toxicity} (Fig.~\ref{fig:toxicity}) exhibit the most significant decreases compared to other attributes.
Moreover, we observe that the Twitter pre-intervention time series exhibits much more variability compared to the post-intervention period. We found that this variability was largely due to the difference in the amount of data collected: 24.16 million posts in the pre-policy timeline versus 9.69 million posts in the post-policy timeline dataset. Additionally, we identified fewer posts that were toxic (i.e., above the 0.5 threshold) in the post-policy dataset. We also note a significant peak for all the perspective attributes, except for \emph{Profanity}, around days 600–620. This spike was attributed to the appointment of Jack Smith as the special counsel in the investigations involving Former President Donald Trump~\cite{jsmith-1,jsmith-2}.  

\begin{table}[t]
\centering

            \caption{DiD regression results for toxicity attributes.}
            \resizebox{\columnwidth}{!}{%
            \begin{tabular}{lll}
             \hline  \hline       
             Event &\textit{Dependent variable: Toxicity}&Confidence Intervals \\
             \hline          
           Treatment & 0.1058 (0.000)$^{***}$ & [0.099, 0.113]\\         
           Post Treatment & -0.0028 (0.411)& [-0.010, 0.004] \\ 
           DiD ($\delta$) & -0.0814 (0.000)$^{***}$& [-0.091, -0.072]  \\
            \hline
            \hline
            Event &\textit{Dependent variable: Severe Toxicity}&Confidence Intervals  \\
             \hline          
           Treatment& 0.1173 (0.000)$^{***}$& [0.114, 0.120] \\         
            Post Treatment & -0.0024 (0.099)&[-0.005, 0.000] \\ 
           DiD ($\delta$) & -0.0570 (0.000)$^{***}$&[-0.061, -0.053] \\
           \hline
            \hline
            Event &\textit{Dependent variable: Profanity}&Confidence Intervals  \\
             \hline      
            Treatment& 0.0691 (0.000)$^{***}$&[0.063, 0.075] \\         
            Post Treatment & 0.0012 (0.702)&[-0.005, 0.007] \\ 
           DiD ($\delta$) & -0.0693 (0.000)$^{***}$&[-0.078, -0.061] \\
           \hline
            \hline
            Event &\textit{Dependent variable: Threat}&Confidence Intervals \\
             \hline      
        Treatment& 0.1605 (0.000)$^{***}$&[0.157, 0.164] \\         
            Post Treatment & -0.0044 (0.025)$^{*}$&[-0.008, -0.001] \\ 
           DiD ($\delta$) & -0.0414 (0.000)$^{***}$&[-0.047, -0.036] \\
           \hline
            \hline
            Event &\textit{Dependent variable: Insult}&Confidence Intervals  \\
             \hline      
        Treatment& 0.1479 (0.000)$^{***}$&[0.143, 0.153] \\         
            Post Treatment & 0.0082 (0.000)$^{***}$&[0.004, 0.013] \\ 
           DiD ($\delta$) & -0.0820 (0.000)$^{***}$&[-0.089, -0.075] \\
           \hline
            \hline
            Event &\textit{Dependent variable: Identity Attack}&Confidence Intervals \\
             \hline      
        Treatment& 0.1282 (0.000)$^{***}$&[0.125, 0.131] \\         
            Post Treatment & 0.0087 (0.000)$^{***}$&[0.006, 0.012] \\ 
           DiD ($\delta$) & -0.0382 (0.000)$^{***}$&[-0.042, -0.034] \\
           \hline
           \textit{Note:}  & \multicolumn{1}{l}{$^{*}$p$<$0.05; $^{**}$p$<$0.01; $^{***}$p$<$0.001} \\ \hline
            \end{tabular}
            }
            \label{stat-linearits}
\end{table}

\textbf{Summary: } 
In summary, our DiD model revealed a statistically significant causal association between Parler's stricter moderation guidelines and a decrease across all toxicity  attributes. We observed that the \emph{Treatment} variable (i.e., $\beta_1$) was positive for all  attributes, indicating that, on average, Parler users posted more abusive content than Twitter users before the intervention. Additionally, \textit{Threat} was the only variable that showed a statistically significant decrease in both Parler and Twitter.
Thus, we can conclude that Parler's changes to its content moderation policies had a positive causal effect in decreasing the toxic and abusive content produced by its users, which directly addresses our RQ1.

\subsection{User Characteristics and Content}
\begin{table} [t]
            \centering
            \caption{Comparison of Users Characteristics} 
            \resizebox{\columnwidth}{!}{%
            \begin{tabular}{l|cccc|cccc}  
            \hline \hline
             &
             \multicolumn{4}{c|}{Pre Policy Change} &
             \multicolumn{4}{c}{Post Policy Change}\\
              &Min &Max &Mean &Median &Min &Max &Mean &Median \\
             \hline
             
            Followers &0 &2,300,000 &20.65 &1 &0 &6,048,750 &34.8 &1\\
            
            Following &0 &126,000 &28.28 &6  &0 &479,412 &33.4 &8\\
            
            \hline
            \end{tabular}
            }
            \label{metrics-user}
\end{table}
\subsubsection{Comparing Users' Characteristics in Pre- and Post-Policy Change Datasets} 
We employed the Mann-Whitney test and could reject the null hypothesis that users in the pre and post-policy change datasets have the same distribution for \emph{followers} and \emph{followings}.
We found an increase in the number of \emph{followings} ($Med_{pre}=6$ vs. $Med_{post}=8$)
and \textit{followers} ($Mean_{pre}=20.65$ vs. $Mean_{post}=34.8$), $p<0.0001$, hence indicating that users are still active on Parler.

\begin{table*}[h!]
    \centering 
    \caption{Badges assigned to users in the pre and post-policy policy change datasets}
    \resizebox{\textwidth}{!}{%
    \begin{tabular}{l|p{0.9\textwidth}|c|c}
        \hline 
        Badge & Description & Pre Change  & Post Change \\
        \hline 
    
        Verified & This badge means Parler has verified the account belongs to a real person and not a bot. Since verified users can change their screen name, the badge does not guarantee one's identity.& 25,734 &  236,431 \\
        Gold & A Gold Badge means Parler has verified the identity of the person or organization. Gold Badges can be influencers, public figures, journalists, media outlets, public officials, government entities, businesses, or organizations (including nonprofits). If the account has a Gold Badge, its parleys and comments come from real people. & 589 &  668 \\
        Integration Partner & Used by publishers to import articles and other content from their websites& 64 &  N/A \\
        RSS feed & These accounts automatically post articles directly from an outlet's website& 99 & 13  \\
        Private &If you see this badge, the account owner has chosen to make the account private. This badge may also be applied to accounts that are locked due to community guideline violations & 596,824  & 337,717\\
        Verified Comments & Users with a verified badge who are restricting comments to only other verified users. & 4,147  & N/A \\
        Parody & Parler approved parody accounts.& 37 & N/A\\
        Parler Employee & This badge is applied to Parler employees' personal accounts, should they wish. Their parleys are their own views and not Parler's.& 25  & 28\\
        Real Name & Users using their real name & 2  & N/A\\
        Parler Early & Signifying Parler's earliest members, this badge appears on accounts opened in 2018.& 81  &  822\\
        Parler Official & These accounts - @Parler, @ParlerDev, and others - issue official statements from the Parler team.& N/A&5\\
    
        \hline
\end{tabular}
}
\label{badges-total}
\end{table*}
Table~\ref{badges-total} presents the number of badges assigned to users in the pre- and post-moderation policy datasets.
We observed a significant increase in the number of users undergoing Parler’s verification process to confirm that their accounts were not bots. This sharp rise in verified users may reflect concerns about an influx of bots as Parler expanded.
Additionally, we found an increase in the number of \emph{Gold} badges, suggesting that some users gained enough popularity post-moderation changes to qualify for this status. These increases indicate that users remained active on the platform following the policy changes.
Interestingly, the number of users with the \emph{Private} badge decreased in the post-moderation dataset. Notably, we did not attempt to collect parleys from users with the \emph{Private} badge; rather, badge information was extracted from user metadata.

\subsubsection{Content Analysis} 
We observed significant interest in the 2020 U.S. elections in the pre-policy change dataset, because the elections took place during the data collection period~\cite{aliapoulios2021large}.
Additionally, we noticed a decline in the usage of phrases like Where We Go One, We Go All (WWG1WGA), which is associated with the QAnon conspiracy movement. Parler-specific terms, such as Parleys, were also more prevalent in the earlier dataset.
Conversely, we observed an increased use of the word patriots, a term Republican lawmakers used to describe the rioters~\cite{patroit}.

\begin{table}[h!]
    \centering 
    \caption{Most Popular Websites Shared on Parler}
    \resizebox{\columnwidth}{!}{%
    \begin{tabular}{l|c|c|c}
        \hline
        \hline 
        Website & Pre Policy & Post Policy & Change(\%)\\
        \hline 
        
        image-cdn.parler.com & 7,318,992 &  1&$-$99.99\\
        youtube.com & 2,499,198 & 225,562 & $-$83.44 \\
        youtu.be & 1,812,871 & 19 & $-$99.99 \\
        bit.ly & 893,603 & 5& $-$99.99\\
        twitter.com & 803,514 & 42,638& $-$89.92\\ 
        media.giphy.com & 539,389 & 545 & $-$99.79 \\ 
        i.imgur.com & 532,365 & 5,779 & $-$97.85 \\
        facebook.com & 520,796 & 318& $-$99.87 \\ 
        thegatewaypundit.com & 469,855 & 610,512  & $+$13.01\\ 
        breitbart.com & 328,953 & 240,547 &$-$15.52 \\
        foxnews.com & 298,285 & 136,956 &$-$37.06 \\
        instagram.com & 168,160 & 22,932 &$-$75.99\\
        rumble.com & 164,949 & 744,132 &$+$63.71 \\
        theepochtimes.com & 136,294 & 33,937 &$-$60.12 \\ 
        hannity.com &13,017 & 148,026 &$+$83.83\\
        justthenews.com & 50,638& 147,984 &$+$49.01 \\
        www.theblaze.com & 2,006& 122,111 &$+$96.76 \\
        www.westernjournal.com & 6,399& 119,551 &$+$89.83 \\
        bongino.com & 17,251 & 114,334 & $+$73.77\\
        www.bitchute.com & 104,462& 87,672 &$-$8.73 \\
        \hline
    \end{tabular}
}
\label{top_websites}
\end{table}

\subsubsection{Links Shared in Parleys} 
From Table~\ref{top_websites}, we observe a sharp rise in the popularity of Rumble links (64\%). This increase is likely due to Rumble's stance on not removing content related to misinformation and election integrity, with MBFC labeling the website as \emph{Right Biased and Questionable}~\cite{singhal2022sok,mbfcrumble}.
In contrast, we observed a decline in the number of Twitter links being shared on Parler. These trends may be attributed to the increasing rhetoric surrounding censorship on Twitter and other popular social media platforms~\cite{Vogels_2020}.
We also observed a sharp increase in the number of \textit{The Blaze links} (97\%) being shared. According to the MBFC service, this website is labeled as \textit{Strongly Right Biased and Questionable}~\cite{mbfclabelssss}. 

Furthermore, we analyzed the links shared on Parler using the MBFC service. We were able to collect labels for 3,937 (2.59\%) and 1,081 (1.75\%) of the total links shared on Parleys from the pre- and post-moderation policy change datasets, respectively. However, we could not collect labels for all URLs, as many were from websites, such as YouTube, Twitter, and Instagram, for which MBFC does not provide labels (see Table~\ref{top_websites}). Despite this, our results are still generalizable, as we were able to capture the majority of websites for which MBFC does provide labels, allowing us to assess the impact of the policy change on users' speech. 

\begin{figure}[h]
    \centering 
    \subfloat[Histogram of Factuality Scores]{\includegraphics[width=0.24\textwidth]{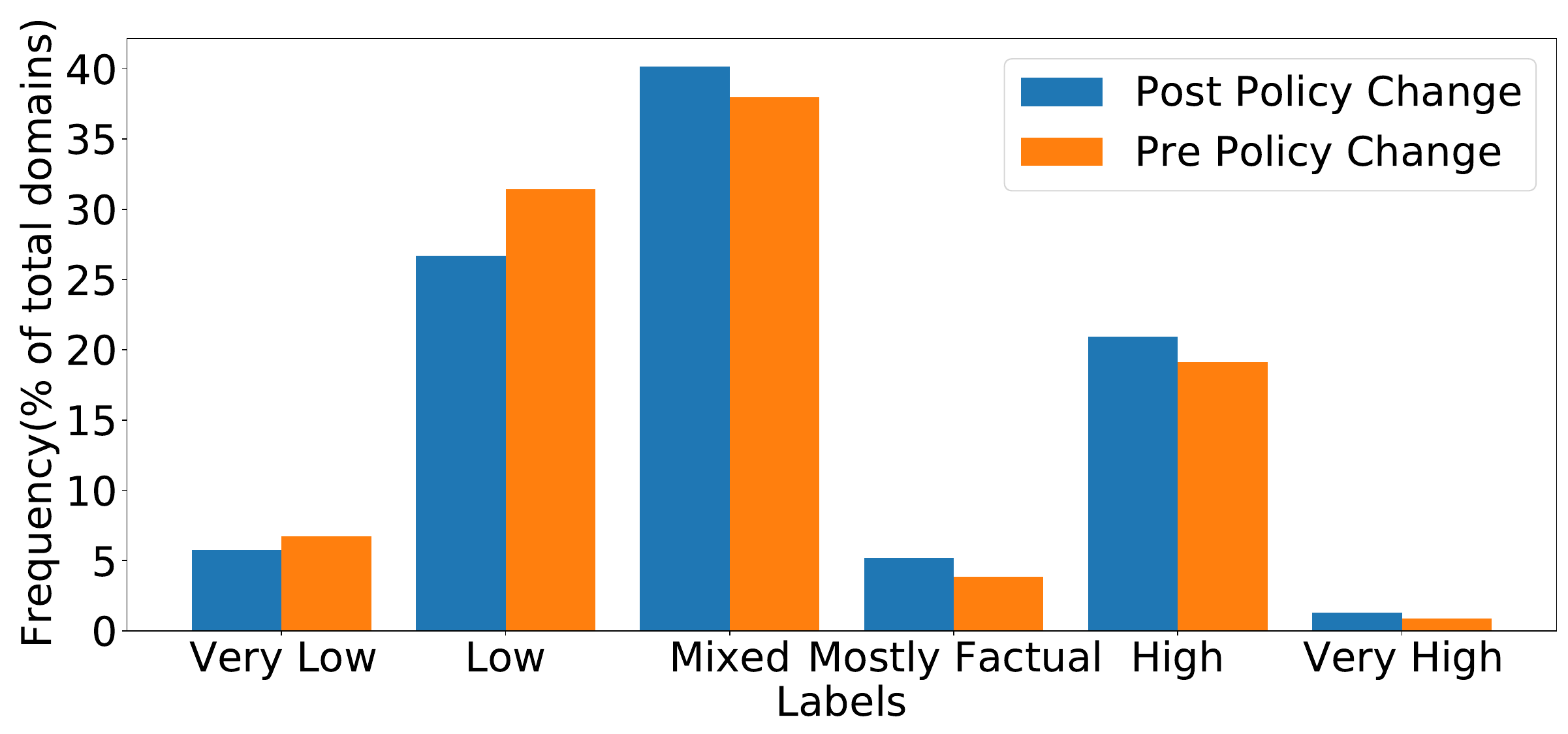} \label{fig:factuality}}
    \subfloat[Histogram of Bias]{\includegraphics[width=0.24\textwidth]{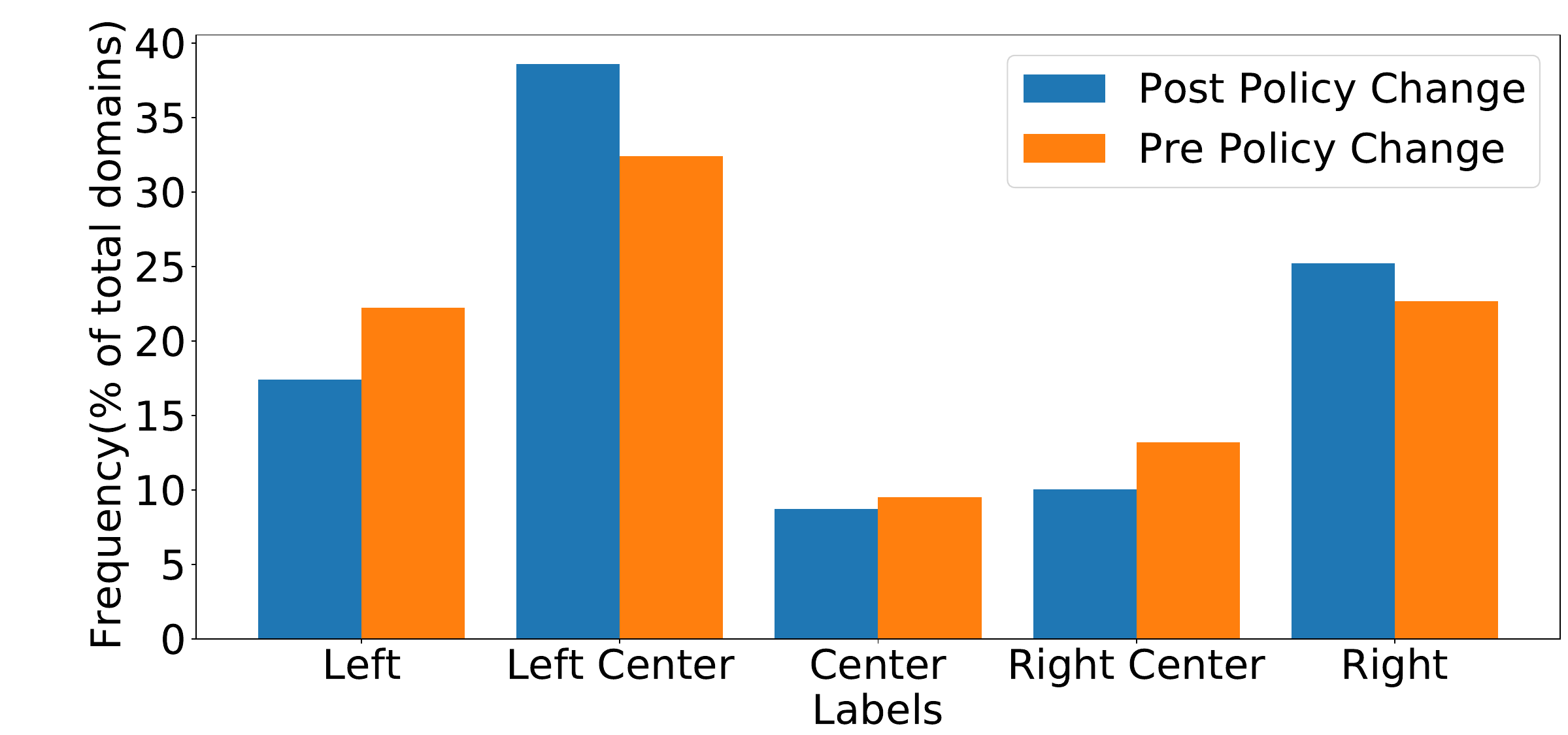} \label{fig:bias}} \\
    \subfloat[Histogram of \\ Conspiracy-Pseudoscience] {\includegraphics[width=0.24\textwidth]{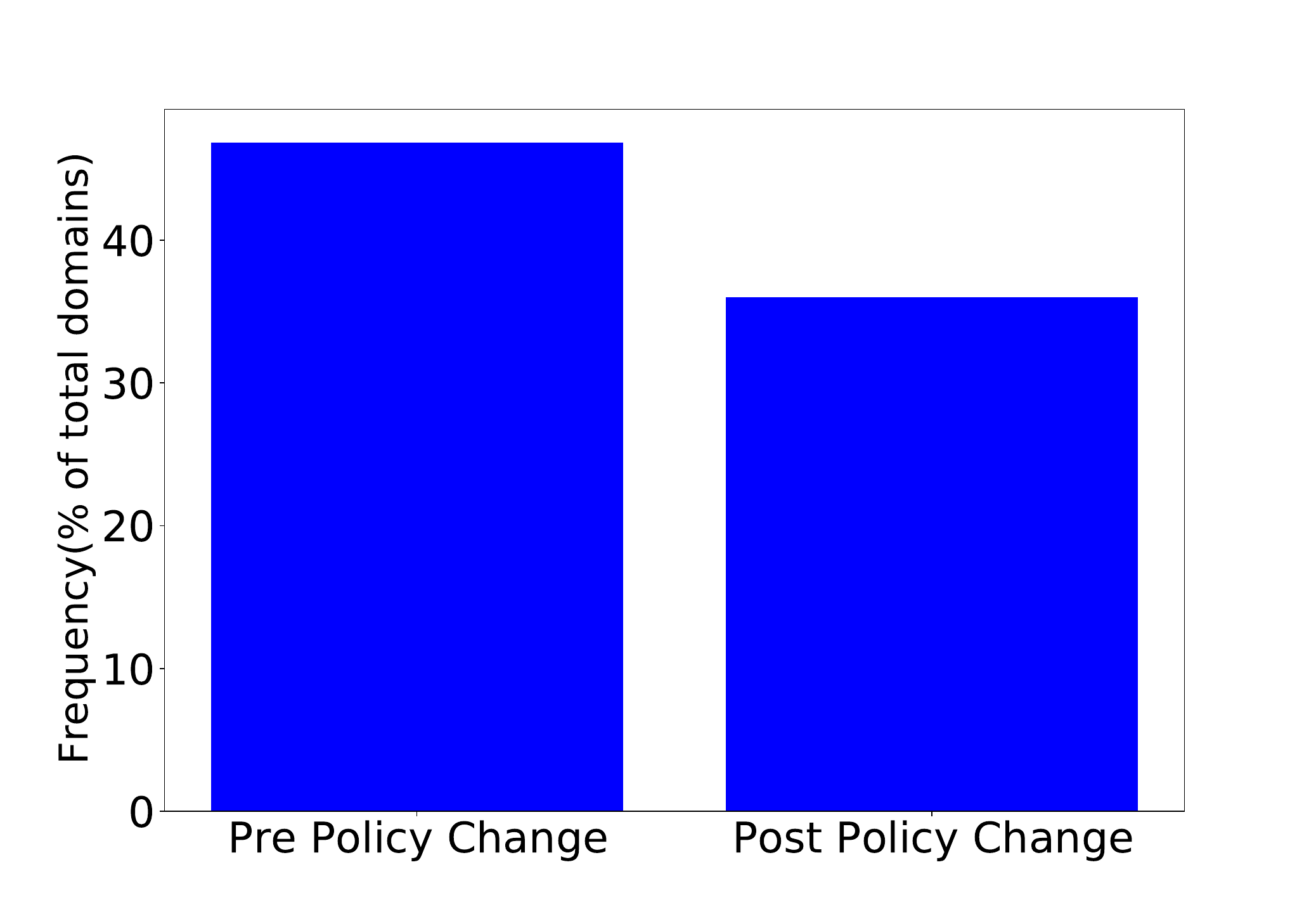} \label{fig:conspiracy}}
    \subfloat[Histogram of Questionable \\ Sources]{\includegraphics[width=0.24\textwidth]{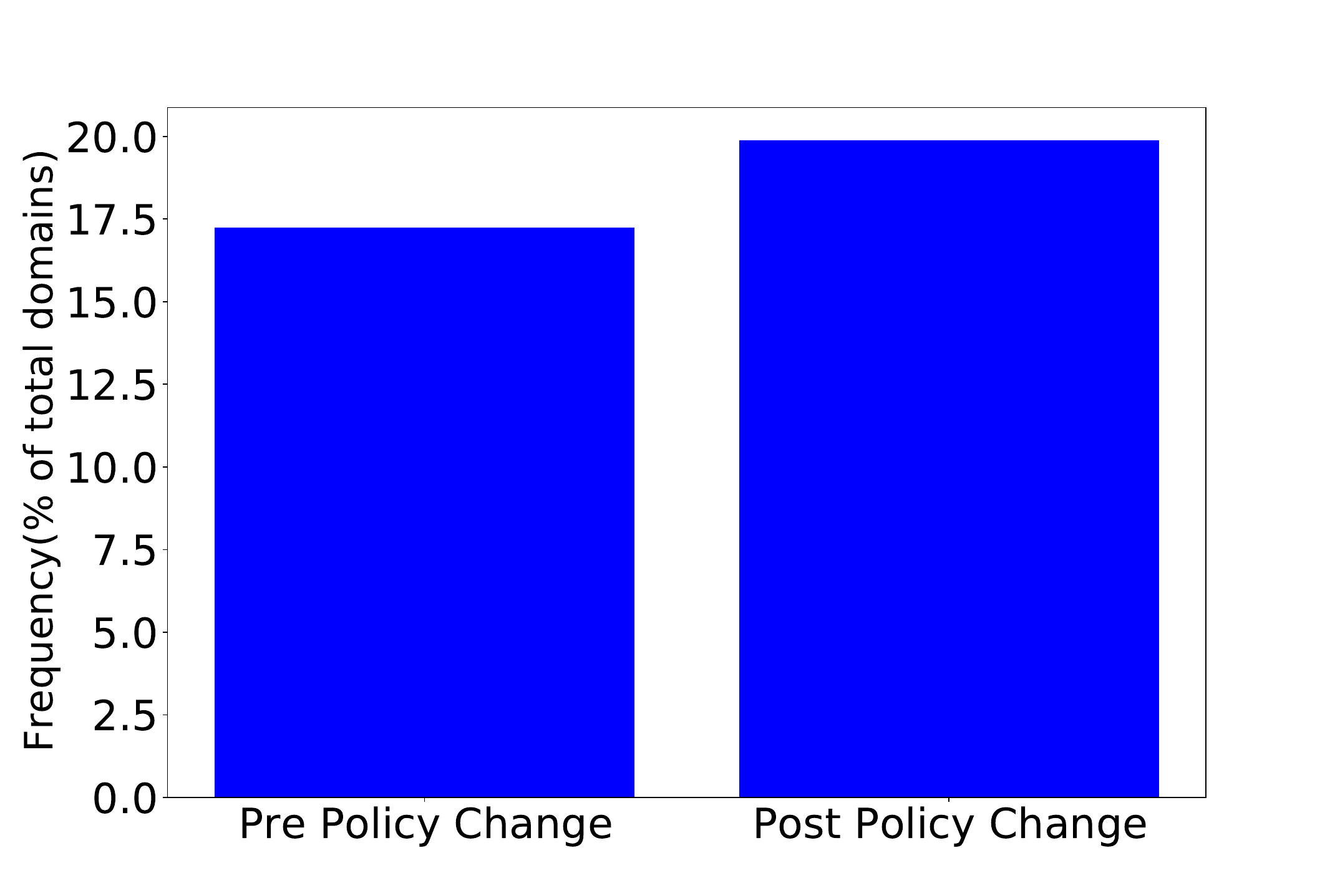} \label{fig:qsource}} 
    \caption{Histogram of MBFC Labels} 
    \label{persssss}
\end{figure}

Figure~\ref{persssss} presents our results. We observe a decrease in the number of conspiracy-pseudoscience news articles, as shown in Figure~\ref{fig:conspiracy}. However, interestingly, there is an increase in the number of \emph{questionable source} articles being shared in the post-policy change dataset, as depicted in Figure~\ref{fig:qsource}. 
This suggests that Parler continues to allow URLs spreading overt propaganda and fake news, aligning with the findings of~\cite{singhal2022sok}. 
In Figure~\ref{fig:factuality}, we observe that most links with a score between \emph{Very Low} and \emph{Low} are from the pre-policy change dataset, while post-moderation links are more evenly distributed across higher ranges, from \emph{Low} to \emph{High}. 
Interestingly, in Figure~\ref{fig:bias}, we observe that Parler users are sharing more URLs from \emph{Left Center} and \emph{Right} websites. This is notable, considering that the majority of Parler users are highly conservative~\cite{collins_2021}. 
In summary, using the labels provided by MBFC, we found that the credibility (factuality) of the URLs being shared did increase. Additionally, there was a substantial decrease in the number of conspiracy-pseudoscience news articles. This is particularly interesting, as in 2022, notable conspiracy theories circulated, such as the claims that mpox (monkeypox) was orchestrated by vaccine manufacturers, that Bill Gates was involved in the outbreak, that it was transmitted solely via sexual interactions, and that the WHO released the virus to gain more power~\cite{zenone2022using,asd}. However, despite these shifts, Parler users were now sharing more URLs from questionable sources than before.

\textbf{Summary:} We observed a statistically significant increase in the number of followings after Parler came back online. Additionally, we found that Parler users were more likely to have their accounts \emph{Verified} compared to the pre-moderation change dataset. Interestingly, the term \emph{patriots} became increasingly prevalent in users' posts. We found an increase in the credibility (factuality) of the URLs being shared. Moreover, we noticed a substantial decrease in the number of conspiracy and pseudoscience news articles. However, Parler users were sharing more questionable source URLs post-moderation than before. These findings indicate substantial changes in Parler users' behavior, effectively answering our RQ2.

\section{Discussion and Future Work}
Our results indicate a positive impact of the changes to Parler's content moderation guidelines following its ban. Our quasi-experimental analysis revealed that, after the policy changes, all Perspective attributes experienced a statistically significant decrease ($p<0.001$). As shown in Table~\ref{stat-linearits}, we observed that \emph{Severe Toxicity}, \emph{Threat}, and \emph{Identity Attack} saw the largest decreases compared to other attributes. This is particularly interesting as it contrasts with findings from prior studies, which observed an increase in toxic rhetoric among users. In contrast, our research highlights that when Parler adjusted its guidelines, the toxic rhetoric from existing users actually decreased. Additionally, using MBFC, we found that the \emph{credibility} (factuality) of URLs shared by Parler users increased, which directly contradicts the observations made in~\cite{trujillo2022make}.

\textbf{Effectiveness of moderation policy.} 
Our results demonstrate that stricter content moderation policies can significantly reduce the toxic rhetoric of existing users. However, an important consideration is the migration of some users from Parler to other fringe social media platforms, such as Rumble, Gab, and Telegram. Studies have shown that users often become more active on fringe platforms after being deplatformed~\cite{horta2023deplatforming}, underscoring the need for tailored moderation strategies~\cite{cresci2022personalized,singhal2022sok}.
Moreover, the migration to fringe platforms can have unintended consequences, with some users exhibiting even more extreme or toxic behavior in these spaces~\cite{horta2021platform}. This raises the question of how content moderation can be effectively managed across multiple platforms, given the risk of users shifting to less regulated environments. Future research should focus on developing coordinated, cross-platform moderation strategies that not only target harmful behavior within individual platforms but also consider the broader ecosystem of social media.

\textbf{Importance of this study.} To the best of our knowledge, this paper is the first study to assess the effectiveness of platform-wide policy changes on all active users present during both the pre- and post-policy periods. This inclusive analysis provides a comprehensive view of the broader ecosystem, as opposed to a limited focus on deplatformed users or specific audiences. Second, while prior research primarily examines hard content moderation measures (e.g., suspensions or removals), our study focuses on replatforming— a unique scenario involving the reinstatement of a platform under a series of progressive policy changes. Third, evaluating the effectiveness of platform-wide moderation is typically challenging due to limited platform transparency and access to comprehensive datasets. However, the unique context of Parler's replatforming allowed us to use a custom crawler to collect data from all active users post-replatforming and to conduct robust comparisons of platform-wide activity and content trends.

Moreover, this study provides the first-ever Parler dataset following its replatforming, along with a framework that can be used to gather data from Parler. This dataset presents a unique opportunity for researchers to examine user behavior, interactions, and the topics discussed within a specific group of users who share particular ideological or political leanings. It offers valuable insights into the dynamics of online communities with more defined mindsets, helping to deepen our understanding of how content moderation and platform policies influence user engagement and discourse. 

Furthermore, Parler was acquired by Starboard in 2023 and temporarily shut down on the same day~\cite{parlershut}. However, the platform has since returned online, rebranded as Parler 3.0~\cite{parler333}, and implemented significant changes to its content moderation guidelines~\cite{parler-newguide}. As a result, both our dataset and the accompanying framework offer a unique opportunity for researchers to assess the evolution of Parler's policy changes, from its inception in 2018 (Parler 1.0), to the first major policy overhaul in 2021 (Parler 2.0), and now to the current iteration, Parler 3.0. 
Additionally, Singhal et al.\cite{singhal2022sok} previously found that Parler lacked any form of soft moderation. However, with the introduction of Time Out—a new soft moderation tool under Parler 3.0's updated guidelines\cite{parler-newguide}—our dataset provides an ideal resource for studying the effectiveness of this intervention in moderating user behavior.

\textbf{Limitations}
In our current dataset, i.e., the post-policy change dataset, we were unable to collect a random sample of users, which limits the generalizability of our analysis and may not fully capture the complete impact of the moderation policy change. Another limitation of our study is that some users may have changed their usernames when Parler was reinstated, possibly to evade detection. Additionally, we acknowledge that Google's Perspective API, used for toxicity detection, has certain limitations and biases~\cite{nogara2023toxic, teblunthuis2023misclassification, sap-etal-2022-annotators}. Furthermore, our work does not account for the impact of users' hateful rhetoric when they migrated to other platforms after Parler was taken offline.
In future research, we plan to investigate user comments on posts to assess whether the moderation changes are reflected in these interactions. Comments may provide additional insights into the effects of the moderation changes Parler implemented.

\section{Conclusion}
On January 12, 2021, Parler was removed from the Apple and Google App Stores, and Amazon Web Services stopped hosting Parler’s content shortly thereafter. This action was attributed to Parler’s refusal to remove posts inciting violence following the 2021 U.S. Capitol riots. Parler was later reinstated after strengthening its moderation policies to address hateful content. Our study investigates the impact of these policy changes on user discourse by comparing user rhetoric in pre- and post-policy change datasets.
Our quasi-experimental analysis shows that, following the moderation changes, all forms of toxicity experienced a significant decrease ($p<0.001$). Additionally, we observed an increase in the factuality of the news sites being shared, along with a decrease in the number of conspiracy or pseudoscience sources being shared.

\begin{acks}
This paper is based upon work supported by NSF CNS 2309318 award. Any opinions, findings, conclusions or recommendations expressed in this material are those of the author(s) and do not necessarily reflect the views of the National Science Foundation.
\end{acks}

\bibliographystyle{ACM-Reference-Format}
\balance
\bibliography{refs}

\end{document}